\documentclass{kluwer}    

\usepackage{graphicx}

\newdisplay{guess}{Conjecture}

\begin{document}
\begin{article}
\begin{opening}         
\title{Exchange Processes in a Multi-Phase ISM} 
\author{Stefan \surname{Harfst}\footnote{harfst@astrophysik.uni-kiel.de}}
\author{Christian \surname{Theis}}
\author{Gerhard \surname{Hensler}}
\runningauthor{S. Harfst et al.}
\runningtitle{Exchange Processes in a Multi-Phase ISM}
\institute{Institute for Theoretical Physics and Astrophysics, University Kiel, Germany}
\date{November 27, 2001}

\begin{abstract}
We present a new particle based code with a multi-phase description of the 
ISM implemented in order to follow the chemo-dynamical evolution of 
galaxies. The multi-phase ISM consists of clouds (sticky particles) and 
diffuse gas (SPH):
Exchange of matter, energy and momentum is achieved by drag (due to ram 
pressure) and condensation or evaporation. Based on time scales we 
show that in Milky-Way-like galaxies the drag force is for molecular clouds 
only important, if their relative velocities exceed 100 km/s. For the
mass exchange we find that clouds evaporate only if the temperature of the 
ambient gas is higher than one million Kelvin. At lower temperatures 
condensation takes place at time scales of the order of 1--10~Gyr.
\end{abstract}
\keywords{Methods: N-body simulations, Galaxies: evolution, Galaxies: ISM}
\end{opening}           

\section{On time scales of the different processes}  

So far in 3d-models of galaxies the ISM is mostly described either as a 
diffuse phase with smoothed particle hydrodynamics (SPH) 
(e.g.\ \opencite{HK89}) or as a clumpy phase with sticky particles (SP) 
(e.g.\ \opencite{TH93}). Alternatively, in previous chemo-dynamical models 
a multi-phase ISM is used, but these models are usually restricted to 
spherical or axisymmetric systems (e.g.\ \opencite{TBH92}; \opencite{SHT97}).
In order to extend the chemo-dynamical models to three dimensions we combine 
both treatments in a new particle based code (see also \inlinecite{B01}): 
The warm diffuse gas phase is described by a SPH formalism, whereas the cold 
molecular clouds are represented by a SP scheme. The coupling between the 
gaseous phases is achieved by the following processes: 1) clouds can 
condensate or evaporate (C/E), 2) a drag is exerted by ram pressure and 3) 
cooling can lead to cloud formation due to thermal instabilities.

\begin{figure}[t]
\tabcapfont
\centerline{
\begin{tabular}{p{5cm}@{\hspace{2pc}}p{5cm}}
   \includegraphics[width=5cm]{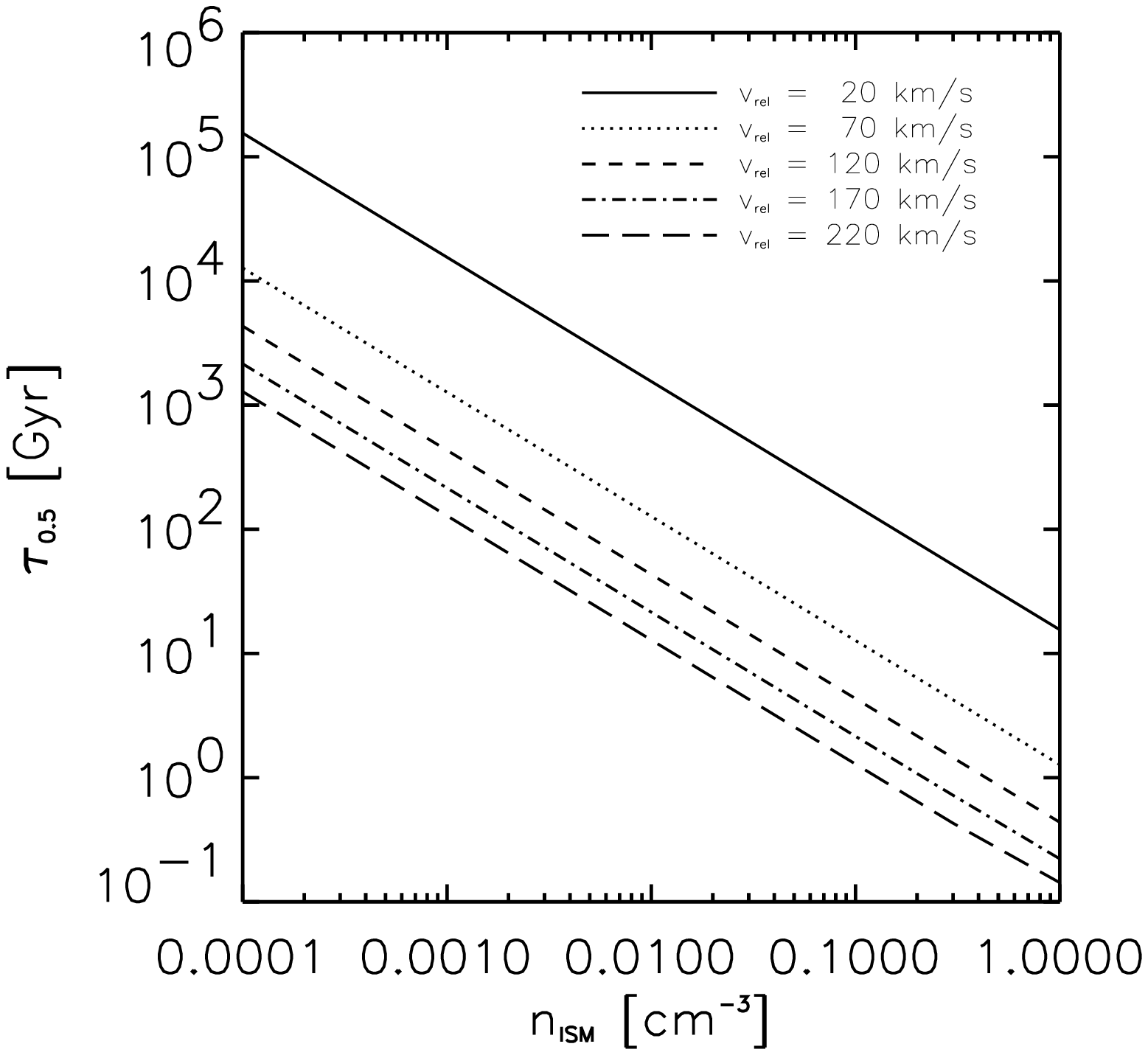} &
   \includegraphics[width=5cm]{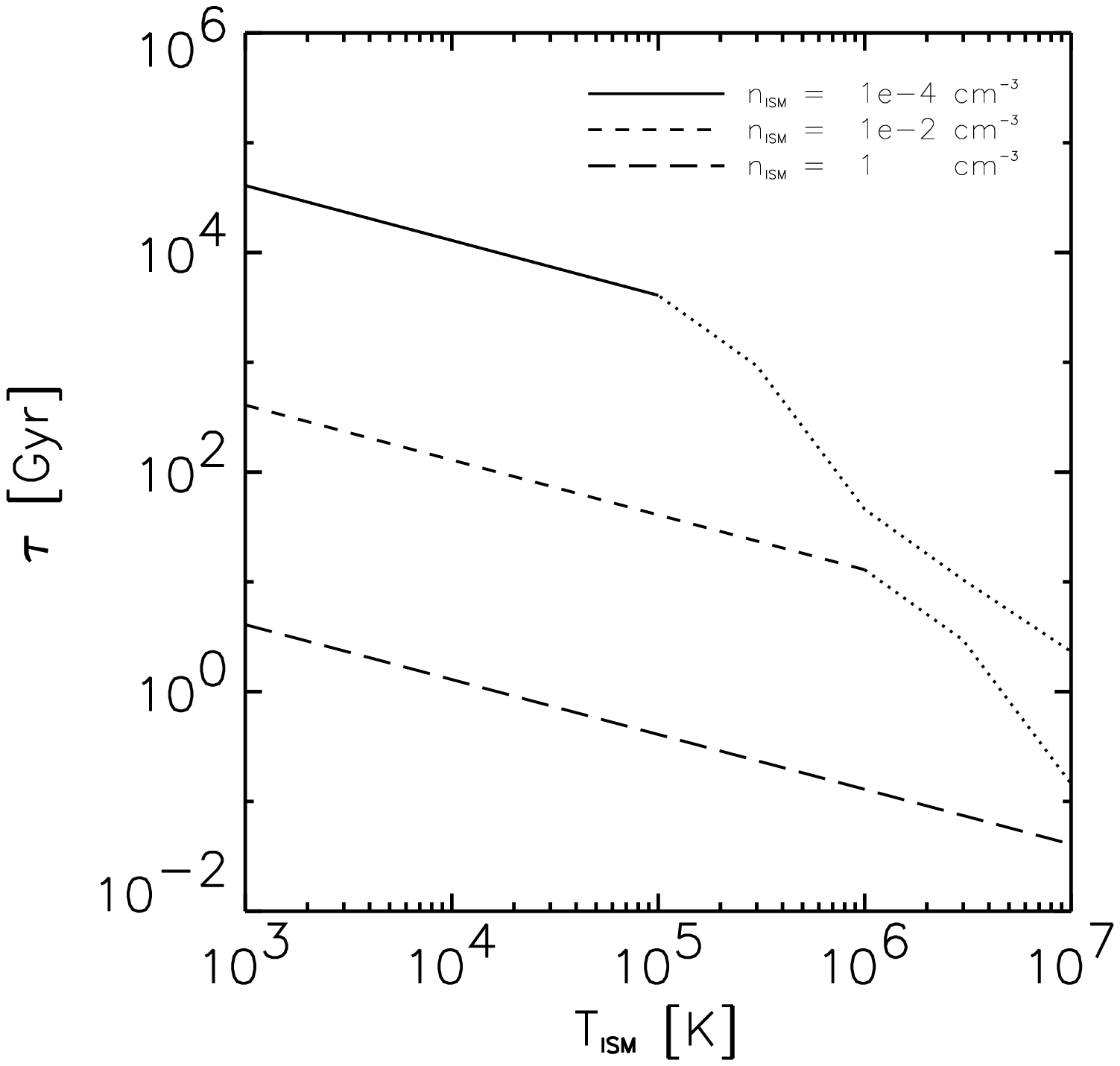} \\
\end{tabular}} 
\caption{The left diagram shows the time scales for drag, which are 
calculated for different densities $n_{\rm ISM}$ and different relative 
velocities $v_{\rm rel}$ of the cloud and the surrounding ISM. $\tau_{0.5}$ 
is defined by the time, a clouds needs to fall to half its initial radius. 
The time scales for C/E (right) are given by 
$\tau = M_{\rm cl} / \dot{M}_{\rm cl}$, i.e.\ the time a clouds need to 
evaporate fully (marked by the dotted lines) or the time to double its mass 
in the case of condensation (marked by the full or dashed lines).
}
\label{fig_ts}
\end{figure}

In the following the influence of drag and C/E on the evolution of disk 
galaxies is discussed in terms of time scales. The drag force and the mass
exchange rates for C/E are calculated following \inlinecite{CMO81} using a
mass-radius-relation based on observations (e.g.\ \opencite{RS87}).
Time scales are determined for different gas densities, temperatures and 
relative velocities (Fig. \ref{fig_ts}):
For low densities the drag force is not important 
($\tau_{\rm drag} > 100\,{\rm Gyr}$). At higher densities and for relative 
velocities greater than $100\,{\rm km}\,{\rm s}^{-1}$ (high velocity clouds) 
$\tau_{\rm drag}$ is of the order of $1\,{\rm Gyr}$ or even less. 
The resulting time scales for C/E range from 
$1-10\,{\rm Gyr}$ for a warm, diffuse gas phase 
($T_{\rm ISM} \approx 10^4\,{\rm K}$ and 
$n_{\rm ISM} \approx 1\,{\rm cm}^{-3}$) up to $100\,{\rm Gyr}$ for hot gas 
component ($T_{\rm ISM} \approx 10^6\,{\rm K}$ and 
$n_{\rm ISM} \approx 10^{-4}\,{\rm cm}^{-3}$). Furthermore, clouds only 
evaporate at temperatures higher than $10^6\,{\rm K}$. This indicates that 
C/E is not very important for the mass evolution of a single cloud, but since
most clouds will accrete matter by condensation, it is possible to mix metals 
from the metal-rich hot gas into the metal-poor clouds.

\vspace{-0.5cm}
\acknowledgements
This work is supported by the {\it Deutsche Forsch\-ungs\-gemein\-schaft (DFG)} 
under the grant TH-511/2-1. The authors would also like to thank the European 
Commission for financial support.

\end{article}
\end{document}